\documentclass[aps,preprint,nofootinbib]{revtex4}
\usepackage{graphics}
\usepackage{slashed}
\usepackage{amssymb}
\usepackage{lscape}
\usepackage{amsmath}
\usepackage{rotating}
\usepackage{bm}
\usepackage{graphicx}
\graphicspath{ {images/} }
\begin{document}
\title{Emission of primordial bosonic radiation during inflation}
\author{$^{2}$ Marcos R. A. Arcod\'{\i}a\footnote{E-mail address: marcodia@mdp.edu.ar},  $^{2}$ Luis Santiago Ridao\footnote{E-mail address: santiagoridao@hotmail.com}, $^{1,2}$ Mauricio Bellini
\footnote{E-mail address: mbellini@mdp.edu.ar} }
\address{$^1$ Departamento de F\'isica, Facultad de Ciencias Exactas y
Naturales, Universidad Nacional de Mar del Plata, Funes 3350, C.P.
7600, Mar del Plata, Argentina.\\
$^2$ Instituto de Investigaciones F\'{\i}sicas de Mar del Plata (IFIMAR), \\
Consejo Nacional de Investigaciones Cient\'ificas y T\'ecnicas
(CONICET), Mar del Plata, Argentina.}
\begin{abstract}
We study the emission of neutral massless $(1, 2)\hbar$-spin bosons during power-law inflation using unified spinor field theory. We shows that
during inflation gravitons and photons were emitted with wavelengths (on physical coordinates)  that increase as the Hubble radius: $\lambda_{Ph} \sim a/H$. The quantised action related to these bosons is calculated and results to be a fraction of the Planck constant.
\end{abstract}
\maketitle

\section{INTRODUCTION}

It is known that the early universe suffered a quasi-exponential expansion that is called inflationary epoch\cite{infl,infl2,infl3,infl4}. This expansion washed away the inhomogeneities of the universe at large scales (or cosmological scales). Inflationary cosmology is the current paradigm of the early universe that describes this primordial cosmological epoch. According to this scenario, the almost constant potential energy $V(\phi)$ of a minimally coupled scalar field $\phi$, called inflaton field, leads to accelerated expansion of space-time. This accelerated expansion of space redshifts all initial matter, leaving a vacuum stated behind. Inflation helps explain the observed large-scale smoothness of the universe, as well as the absence of unwanted relics such as magnetic monopoles. Most excitingly, quantum fluctuations during the inflationary period can be amplified to density perturbations\cite{prd96} that seed the formation of galaxies and large-scale structure in the universe. These scales correspond today between $10^8$ and $10^{10}$ light years, but at that time were bigger than the size of the observable universe. Inflation is now supported by many observational evidences, in particular, by
the\ discovery of temperature anisotropies present in the cosmic microwave
background (CMB) \cite{re1,re1p}. In fact, in recent years there has been an
extraordinary development on observational tests of inflationary models
\cite{Bi2}. On the theoretical side, among the most popular and pioneering
models of inflation we would like to mention the supercooled chaotic
inflation\ model \cite{re2}. In this proposal, as we know, the expansion of
the universe is driven by a scalar field known as the inflaton field.

It is expectable that in the inflationary epoch, coherent radiation of bosons with spin $s=\hbar$ and $s=2\hbar$ has been emitted and the possibility that can be detected today becomes more close because of the last advances in the detection of gravitational waves\cite{ligo}. In this work we study
the emission of both kinds of radiation during a power-law inflationary expansion using the recently introduced formalism called Unified Spinor Fields (USF)\cite{MM,SMM}. It is expected that a fraction of the energy density in the universe is given by this kind of coherent radiation, with wavelengths of the order of cosmological scales. They should be a condensate of bosons which should be an important part of dark energy in the universe. In this work we calculate the action due to the both kinds of radiation, and we demonstrate that this action can be quantised.
The paper is organised as follows: In Sect. II we describe the Einstein-Hilbert action and the quantum structure of space-time. In Sect. III we revisited the dynamics of bosons. In Sect. IV we introduce the quantised action. In Sect. V we study power-law inflation and we obtain the frequency of photons and gravitons in physical coordinates during inflation. The quantised action for massless neutral boson fields is calculated. Finally, in Sect. VI we develop some final comments.

\section{EINSTEIN-HILBERT ACTION AND QUANTUM STRUCTURE OF SPACE TIME}\label{ei}

If we deal with an orthogonal basis,
the curvature tensor will be written in terms of the connections:
$R^{\alpha}_{\,\,\,\beta\gamma\delta} = \Gamma^{\alpha}_{\,\,\,\beta\delta,\gamma} -  \Gamma^{\alpha}_{\,\,\,\beta\gamma,\delta}
+ \Gamma^{\epsilon}_{\,\,\,\beta\delta} \Gamma^{\alpha}_{\,\,\,\epsilon\gamma} - \Gamma^{\epsilon}_{\,\,\,\beta\gamma}
\Gamma^{\alpha}_{\,\,\,\epsilon\delta}$). The Einstein-Hilbert (EH) action for an arbitrary matter lagrangian density ${\cal L}$
\begin{equation}
{\cal I} = \int d^4 x \sqrt{-g} \left[ \frac{R}{2\kappa}+ {\cal L} \right],
\end{equation}
after variation, is given by
\begin{equation}\label{delta}
\delta {\cal I} = \int d^4 x \sqrt{-g} \left[ \delta g^{\alpha\beta} \left( G_{\alpha\beta} + \kappa T_{\alpha\beta}\right)
+ g^{\alpha\beta} \delta R_{\alpha\beta} \right],
\end{equation}
where $\kappa = 8 \pi G$, $G$ is the gravitational constant and $g^{\alpha\beta} \delta R_{\alpha\beta} =\nabla_{\alpha}
\delta W^{\alpha}$, such that $\delta W^{\alpha}=\delta
\Gamma^{\alpha}_{\beta\gamma} g^{\beta\gamma}-
\delta\Gamma^{\epsilon}_{\beta\epsilon}
g^{\beta\alpha}$. When the flux of $\delta W^{\alpha}$ that cross the Gaussian-like hypersurface defined on an arbitrary region of the spacetime,
is zero, the resulting equations that minimize the EH action, are the background Einstein equations: $G_{\alpha\beta} + \kappa\, T_{\alpha\beta}=0$. However, when this flux is nonzero,
one obtains in the last term of the eq. (\ref{delta}), that
$\nabla_{\alpha} \delta W^{\alpha}=\delta\Theta(x^{\alpha})$, such that $\delta\Theta(x^{\alpha})$ is an arbitrary scalar field. This flux becomes zero when there are no sources inside this hypersurface. Hence, in order to make
$\delta {\cal I}=0$ in Equation (\ref{delta}), we must consider the condition: $
G_{\alpha\beta} + \kappa T_{\alpha\beta} = \Lambda\,
g_{\alpha\beta}$, where $\Lambda$ is the cosmological constant. Additionally, we must require
the constriction $\delta g_{\alpha\beta} \Lambda =
\delta\Theta\, g_{\alpha\beta}$, in order to be complied the gauge-invariant transformations: $\bar{\delta W}_{\alpha} = \delta W_{\alpha} - \nabla_{\alpha} \delta\Theta$, where the scalar field $\delta\Theta$ complies $\Box \delta\Theta =0$\cite{rb}. On the other hand, we can make the transformation
\begin{equation}\label{ein}
\bar{G}_{\alpha\beta} = {G}_{\alpha\beta} - \Lambda\, g_{\alpha\beta},
\end{equation}
and the transformed Einstein equations with the equation of motion for the transformed gravitational waves, hold
\begin{equation}
\bar{G}_{\alpha\beta} = - \kappa\, {T}_{\alpha\beta}. \label{e1} \\
\end{equation}
The Equation (\ref{e1}) provides us with the Einstein equations with cosmological
constant included. Notice that the scalar field $\delta\Theta(x^{\alpha})$ appears
as a scalar flux of the 4-vector with components $\delta W^{\alpha}$
through the closed hypersurface $\partial{\cal M}$. This arbitrary
hypersurface  must be viewed as a
3D Gaussian-like hypersurface situated in any region of
space-time. Furthermore, since $\delta \Theta(x^{\alpha})\, g_{\alpha\beta} = \Lambda\,\delta
g_{\alpha\beta}$, the existence of the cosmological constant $\Lambda$, is related
to the existence of the Gaussian hypersurface, where $g^{\alpha\beta} \delta R_{\alpha\beta}= \delta \Theta$.

In this work we shall use a recently introduced extended Weylian manifold\cite{MM,SMM} to describe quantum geometric spinor fields
$\hat{\Psi}^{\alpha}$, where the connections are
\begin{equation}\label{ga}
\hat{\Gamma}^{\alpha}_{\beta\gamma} = \left\{ \begin{array}{cc}  \alpha \, \\ \beta \, \gamma  \end{array} \right\}+ \hat{\Psi}^{\alpha}\,g_{\beta\gamma}.
\end{equation}
Here
\begin{equation}\label{uch}
\hat{\delta{\Gamma}}^{\alpha}_{\beta\gamma}=\hat{\Psi}^{\alpha}\,g_{\beta\gamma},
\end{equation}
describes the quantum displacement of the extended Weylian manifold with respect to the classical Riemannian background, which is described by the Levi-Civita symbols in (\ref{ga}).

\subsection{Quantum structure of space-time}

In order to describe the quantum structure of space time we consider a the variation $\delta\hat{X}^{\mu}$ of the quantum operator $\hat{X}^{\mu}$:
 \begin{displaymath}
\hat{X}^{\alpha}(x^{\nu}) = \frac{1}{(2\pi)^{3/2}} \int d^3 k \, \bar{\gamma}^{\alpha} \left[ b_k \, \hat{X}_k(x^{\nu}) + b^{\dagger}_k \, \hat{X}^*_k(x^{\nu})\right],
\end{displaymath}
such that $b^{\dagger}_k$ and $b_k$ are the creation and destruction operators of space-time, such that $\left< B \left| \left[b_k,b^{\dagger}_{k'}\right]\right| B  \right> = \delta^{(3)}(\vec{k}-\vec{k'})$and $\bar{\gamma}^{\alpha}$ are $4\times 4$-matrices that comply with the Clifford algebra. Moreover, we shall define in the analogous manner the variation $\delta\hat{\Phi}^{\mu}$ of the quantum operator $\hat{\Phi}^{\mu}$ that describes the quantum inner space:
\begin{displaymath}
\hat{\Phi}^{\alpha}(\phi^{\nu}) = \frac{1}{(2\pi)^{3/2}} \int d^3 s \, \bar{\gamma}^{\alpha} \,\left[ c_s \, \hat{\Phi}_s(\phi^{\nu}) + c^{\dagger}_s \, \hat{\Phi}^*_s(\phi^{\nu})\right],
\end{displaymath}
such that $c^{\dagger}_s$ and $c_s$ are the creation and destruction operators of the inner space, such that $\left< B \left| \left[c_s,c^{\dagger}_{s'}\right]\right| B  \right> = \delta^{(3)}(\vec{s}-\vec{s'})$. In our case the background quantum state can be represented in a ordinary Fock space in contrast with LQG\cite{aa}, where operators are qualitatively different
from the standard quantization of gauge fields. These operators can be applied to some background  quantum state, and describes a Fock space on an arbitrary Riemannian curved space time $\left|B\right>$, such that they comply with
\begin{equation}\label{dif}
\delta\hat{X}^{\mu}\left|B\right> = dx^{\mu}\left|B\right>, \qquad \delta\hat{\Phi}^{\mu}\left|B\right> = d\phi^{\mu}\left|B\right>,
\end{equation}
where $\phi^{\alpha}$ are the four compact dimensions related to their canonical momentum components $s^{\alpha}$ that describe the spin. The states $\left|B\right>$ do not evolves with time because we shall consider the Heisenberg representation, in which only the operators evolve with time so that the background expectation value of the manifold displacement is null: $\left<B\left|\hat{\delta\Gamma}^{\alpha}_{\beta\gamma}\right|B\right>=0$. In order to describe the effective background space-time, we shall consider the line element
\begin{equation}\label{line}
dl^2 \delta_{BB'}= dS^2 \delta_{BB'} + d\phi^2 \delta_{BB'} = \left<B\right| \hat{\delta X}_{\mu} \hat{\delta X}^{\mu} \left| B'\right> + \left<B\right| \hat{\delta \Phi}_{\mu} \hat{\delta \Phi}^{\mu} \left| B'\right>.
\end{equation}
The variations and differentials of the operators $\hat{X}^{\mu}$ and $\hat{\Phi}^{\mu}$ on the extended Weylian manifold, are given respectively by
\begin{eqnarray}
\delta\hat{X}^{\mu}\left| B\right> &=& \left(\hat{X}^{\mu}\right)_{\|\alpha} dx^{\alpha}\left| B\right>, \qquad \delta\hat{\Phi}^{\mu} \left| B\right>= \left(\hat{\Phi}^{\mu}\right)_{\|\alpha} d\phi^{\alpha}\left| B\right>, \\
d\hat{X}^{\mu} \left| B\right>&=& \left(\hat{X}^{\mu}\right)_{,\alpha} dx^{\alpha}\left| B\right>, \qquad d\hat{\Phi}^{\mu} \left| B\right>= \left(\hat{\Phi}^{\mu}\right)_{,\alpha} d\phi^{\alpha}\left| B\right>,
\end{eqnarray}
with covariant derivatives
\begin{eqnarray}
\left(\hat{X}^{\mu}\right)_{\|\beta}\left| B\right> &=& \left[\nabla_{\beta} \hat{X}^{\mu} + \hat{\Psi}^{\mu} \hat{X}_{\beta} - \left(1-\xi^2\right) \hat{X}^{\mu} \hat{\Psi}_{\beta}\right]\left| B\right>, \\
\left(\hat{\Phi}^{\mu}\right)_{\|\beta}\left| B\right> &=& \left[\nabla_{\beta} \hat{\Phi}^{\mu} + \hat{\Psi}^{\mu} \hat{\Phi}_{\beta} - \left(1-\xi^2\right) \hat{\Phi}^{\mu} \hat{\Psi}_{\beta}\right]\left| B\right>.
\end{eqnarray}

\subsection{Bi-vectorial structure of inner space}

We shall consider the squared of the $\hat{\delta\Phi}$-norm on the bi-vectorial space, and the squared $\hat{\delta X}$-norm on the vectorial space, are
\begin{eqnarray}
\underleftrightarrow{\delta\Phi} \overleftrightarrow{\delta\Phi} &\equiv& \left( \hat{\delta\Phi}_{\mu} \hat{\delta\Phi}_{\nu} \right) \left( \bar{\gamma}^{\mu} \bar{\gamma}^{\nu}\right), \\
\underrightarrow{\delta{X}} \overrightarrow{\delta{X}} & \equiv &  \hat{\delta{X}}_{\alpha} \hat{\delta{X}}^{\alpha}.
\end{eqnarray}
such that $\hat{\Phi}^{\alpha}=\frac{1}{2}\phi \,\bar{\gamma}^{\alpha} $ and $\hat{X}^{\alpha}=\frac{1}{2} x\, \bar{\gamma}^{\alpha} $ are respectively the components of the inner and coordinate spaces. Furthermore,  $\bar{\gamma}_{\mu}$ are the ($4\times 4$) Dirac matrices that generate the vectorial and bi-vectorial structure of the space time
\begin{eqnarray}
\left<B\left|\hat{X}_{\mu} \hat{X}^{\mu} \right|B\right> &=& x^2\, \,\mathbb{I}_{4\times 4}, \\
\left<B\left|\left( \hat{\Phi}_{\mu} \hat{\Phi}_{\nu} \right) \left( \bar{\gamma}^{\mu} \bar{\gamma}^{\nu}\right) \right|B\right>&=&
\left<B\left|\frac{1}{4} \left\{ \hat{\Phi}_{\mu} ,\hat{\Phi}_{\nu} \right\} \left\{ \bar{\gamma}^{\mu} ,\bar{\gamma}^{\nu}\right\} -\frac{1}{4}
\left[ \hat{\Phi}_{\mu}, \hat{\Phi}_{\nu} \right] \left[ \bar{\gamma}^{\mu}, \bar{\gamma}^{\nu}\right] \right|B\right>\nonumber \\
&=&\phi^2 \,\mathbb{I}_{4\times 4}. \nonumber
\end{eqnarray}
The $\bar{\gamma}_{\alpha}= E^{\mu}_{\alpha} \gamma_{\mu}$ matrices which generate the background metric are related by the vielbein $E^{\mu}_{\alpha}$ to basis $\gamma_{\mu}$ in the Minkowsky spacetime (in cartesian coordinates). In this paper we shall consider the Weyl basis, such that:
 $\left\{ \gamma^a,\gamma^b\right\} = 2 \eta^{ab} \mathbb{I}_{4\times 4}$. They comply with the Clifford algebra
\begin{equation}
\bar{\gamma}^{\mu} = \frac{\bf{I}}{3!} \left(\bar{\gamma}^{\mu}\right)^2\,\epsilon^{\mu}_{\,\,\alpha\beta\nu} \bar{\gamma}^{\alpha\beta}  \bar{\gamma}^{\nu} , \qquad \left\{\bar{\gamma}^{\mu}, \bar{\gamma}^{\nu}\right\} =
2 g^{\mu\nu} \,\mathbb{I}_{4\times 4}, \nonumber
\end{equation}
where ${\bf{I}}={\gamma}^{0}{\gamma}^{1}{\gamma}^{2}{\gamma}^{3}$, $\mathbb{I}_{4\times 4}$ is the identity matrix, $\bar{\gamma}^{\alpha\beta}=\frac{1}{2} \left[\bar{\gamma}^{\alpha}, \bar{\gamma}^{\beta}\right]$.
\begin{eqnarray}
&& \gamma^0= \,\left(\begin{array}{ll}  0 & \mathbb{I} \\
\mathbb{I}  &  0 \ \end{array} \right),\qquad
\gamma^1=  \left(\begin{array}{ll} 0 &  -\sigma^1 \\
\sigma^1 & 0  \end{array} \right),  \nonumber \\
&& \gamma^2= \left(\begin{array}{ll} 0 &  -\sigma^2 \\
\sigma^2 & 0  \end{array} \right),  \qquad \gamma^3= \left(\begin{array}{ll} 0 &  -\sigma^3 \\
\sigma^3 & 0  \end{array} \right),\nonumber
\end{eqnarray}
such that the Pauli matrices are
\begin{eqnarray}
&& \sigma^1 = \left(\begin{array}{ll} 0 & 1 \\
1  & 0  \end{array} \right), \quad \sigma^2 = \left(\begin{array}{ll} 0 & -i \\
i  & 0  \end{array} \right), \quad \sigma^3 = \left(\begin{array}{ll} 1 & 0 \\
0  & -1  \end{array} \right). \nonumber
\end{eqnarray}

\section{DYNAMICS OF SPINOR FIELDS REVISITED}

To obtain the relativistic dynamics of the spinor fields $\Psi_{\alpha}$ on the extended Weylian manifold, we must calculate the variation of the extended Ricci tensor $\delta{R}^{\alpha}_{\beta\gamma\alpha}=\delta{R}_{\beta\gamma}$: $\delta{R}_{\beta\gamma} = \left(\delta\Gamma^{\alpha}_{\beta\alpha} \right)_{\| \gamma} - \left(\delta\Gamma^{\alpha}_{\beta\gamma} \right)_{\| \alpha}$, which result to be:
\begin{eqnarray}
\hat{\delta{R}}_{\beta\gamma} &=& \nabla_{\gamma} \hat{\Psi}_{\beta} - 3 \left(1-\frac{\xi^2}{3}\right) g_{\beta\gamma} \left(\hat{\Psi}^{\nu} \hat{\Psi}_{\nu}\right) \nonumber \\
&-&g_{\beta\gamma} \left(\nabla_{\nu} \hat{\Psi}^{\nu}\right) +\left(1-\frac{\xi^2}{3}\right)\hat{\Psi}_{\beta}\hat{\Psi}_{\gamma}=\hat{U}_{\beta\gamma}+\hat{V}_{\beta\gamma},
\end{eqnarray}
$\hat{U}_{\beta\gamma}$, and $\hat{V}_{\beta\gamma}$ being respectively the symmetric and antisymmetric parts of $\hat{\delta{R}}_{\beta\gamma}$\cite{MM}:
\begin{eqnarray}
\hat{U}_{\beta\gamma}&=&\frac{1}{2} \left( \nabla_{\beta} \hat{\Psi}_{\gamma}+\nabla_{\gamma} \hat{\Psi}_{\beta}\right)  - g_{\beta\gamma} \left(\nabla_{\nu} \hat{\Psi}^{\nu}\right) \nonumber \\
&-& 3\left(1-\frac{\xi^2}{3}\right) g_{\beta\gamma} \left(\hat{\Psi}_{\nu} \hat{\Psi}^{\nu}\right)  + 3 \left(1-\frac{\xi^2}{3}\right) \left\{\hat{\Psi}_{\beta},\hat{\Psi}_{\gamma}\right\}, \nonumber \\
\hat{V}_{\beta\gamma} &=& -\frac{1}{2} \left( \nabla_{\beta} \hat{\Psi}_{\gamma}-\nabla_{\gamma} \hat{\Psi}_{\beta}\right)+ \frac{3}{2}\left(1-\frac{\xi^2}{3}\right) \left[\hat{\Psi}_{\beta}, \hat{\Psi}_{\gamma}\right] .\nonumber
\end{eqnarray}
Notice that the symmetric and anti-symmetric parts of $\hat{\delta{R}}_{\beta\gamma}$ are due to the algebra described by the quantum operator components $\hat{\Psi}_{\beta}$.
It is easy to show that $\hat{\delta{R}}^{\alpha}_{\beta\alpha\gamma}=-\hat{\delta{R}}_{\beta\gamma}$, so that this tensor gives us redundant information.

On the other hand, the purely antisymmetric tensor
$\hat{\delta{R}}^{\alpha}_{\alpha\beta\gamma}\equiv \hat{\Sigma}_{\beta\gamma}$, is
\begin{equation}\label{sigma}
\hat{\Sigma}_{\beta\gamma} = \left( \nabla_{\beta} \hat{\Psi}_{\gamma}-\nabla_{\gamma} \hat{\Psi}_{\beta}\right) -\left(1+\xi^2\right) \left[ \hat{\Psi}_{\beta}, \hat{\Psi}_{\gamma} \right].
\end{equation}
It is possible to obtain the varied Einstein tensor on the extended Weylian manifold: $\hat{\delta{G}}_{\beta\gamma}= \hat{U}_{\beta\gamma} - \frac{1}{2} g_{\beta\gamma} \hat{U}$, where $\hat{U}=g^{\alpha\beta} \hat{U}_{\alpha\beta}$:
\begin{eqnarray}
\hat{\delta{G}}_{\beta\gamma}&=&\frac{1}{2} \left( \nabla_{\beta} \hat{\Psi}_{\gamma}+\nabla_{\gamma} \hat{\Psi}_{\beta}\right)+\frac{1}{2} g_{\beta\gamma}\left[ \left(1-\frac{\xi^2}{3}\right)\left(\hat{\Psi}^{\alpha}\hat{\Psi}_{\alpha}\right)\right. \nonumber \\
&+& \left. \left(\nabla_{\nu} \hat{\Psi}^{\nu}\right)\right]
+\frac{1}{2}  \left(1-\frac{\xi^2}{3}\right) \left\{\hat{\Psi}_{\beta}, \hat{\Psi}_{\gamma}\right\} . \label{gg}
\end{eqnarray}
In certain cases, it may be useful to describe separately massless and matter spinor fields. To do it we can make a linear combinations to define the tensors\cite{MM}
$\hat{\cal{N}}_{\beta\gamma} = {1\over 2} \hat{V}_{\beta\gamma} - {1\over 4}\hat{{\Sigma}}_{\beta\gamma} $ and $\hat{\cal{M}}_{\beta\gamma} = {1\over 2} \hat{V}_{\beta\gamma} + {1\over 4} \hat{{\Sigma}}_{\beta\gamma} $
\begin{eqnarray}
\hat{\cal{N}}_{\beta\gamma} & = & -\frac{1}{2} \left( \nabla_{\beta} \hat{\Psi}_{\gamma}-\nabla_{\gamma} \hat{\Psi}_{\beta}\right) + \left[ \hat{\Psi}_{\beta}, \hat{\Psi}_{\gamma}\right], \label{n1}\\
\hat{\cal{M}}_{\beta\gamma} & = & \frac{1}{2} \left(1-\xi^2\right) \left[ \hat{\Psi}_{\beta}, \hat{\Psi}_{\gamma}\right], \label{n2}
\end{eqnarray}
such that the symmetric tensor $\hat{\delta{G}}_{\beta\gamma}$, with the antisymmetric ones $\hat{\cal{N}}_{\beta\gamma}$ and $\hat{\cal{M}}_{\beta\gamma}$, describe all the dynamics of possible spinor fields, which must be conserved on the extended Weylian manifold:
\begin{equation}\label{din}
\left(\hat{\delta{G}}^{\beta\gamma}\right)_{\|\gamma} =0, \quad \left(\hat{\cal{M}}^{\beta\gamma}\right)_{\|\gamma}=0, \quad \left(\hat{\cal{N}}^{\beta\gamma}\right)_{\|\gamma}=0.
\end{equation}
Taking into account the gauge-transformations: $\hat{\bar{\delta G}}_{\alpha\beta}= \hat{\delta G}_{\alpha\beta} - g_{\alpha\beta} \hat{\Lambda}$\cite{MM}, we obtain that
\begin{equation}\label{la}
\hat{\Lambda} = - \frac{3}{4} \left[ \nabla_{\alpha} \hat{\Psi}^{\alpha} + \left(1- \frac{\xi^2}{3}\right) \hat{\Psi}^{\alpha} \hat{\Psi}_{\alpha}\right].
\end{equation}
In this paper we are interested only in the last equation in (\ref{din}). This equation can be developed in term of the spinor fields, which in our case are massless bosons
\begin{eqnarray}
\Box {\hat{\Psi}}^{\alpha}&-&\nabla_{\beta} \left(\nabla^{\alpha} \hat{\Psi}^{\beta} \right) +  2 \left( \nabla_{\beta} \hat{\Psi}^{\alpha}\right) \hat{\Psi}^{\beta} - 2 \left(
\nabla_{\beta} \hat{\Psi}^{\beta} \right) \hat{\Psi}^{\alpha} - \left(\nabla^{\alpha}\hat{\Psi}^{\gamma}\right) \hat{\Psi}_{\gamma} + 2 \hat{\Psi}^{\alpha} \left(
\nabla_{\gamma} \hat{\Psi}^{\gamma} \right)   \nonumber \\
 &+& \left(\nabla^{\gamma}\hat{\Psi}^{\alpha}\right)\hat{\Psi}_{\gamma}- 2 \hat{\Psi}^{\gamma} \left(\nabla_{\gamma} \hat{\Psi}^{\alpha}\right) -  2 \left[ \hat{\Psi}^{\mu},
\hat{\Psi}^{\alpha} \right] \hat{\Psi}_{\mu} =0,  \label{a2}
\end{eqnarray}
where bosons must comply\cite{MM}
\begin{eqnarray}
&& \left< B\left| \left[\hat{\Psi}_{\mu}({\bf x}, {\bf \phi}), \hat{\Psi}_{\nu}({\bf x}', {\bf \phi}') \right]\right|B \right> = \frac{s^2 \,L^2_p}{2 \hbar^2 } \left[\hat\gamma_{\mu} , \hat\gamma_{\nu}\right] \, \sqrt{\frac{\eta}{g}} \,\,\delta^{(4)} \left({\bf x} - {\bf x}'\right) \,\delta^{(4)} \left({\bf \phi} - {\bf \phi}'\right). \label{q2}
\end{eqnarray}
The ratio $\sqrt{\frac{\eta}{g}}$, describes the inverse of the relative volume of the background manifold (with metric $g_{\mu\nu}$), with respect to the Minkowski one (with metric $\eta_{\mu\nu}$). The Fourier expansion for the spinor field $ \hat{\Psi}_{\alpha}$ is
\begin{eqnarray}\label{fou}
 \hat{\Psi}_{\alpha}&=& \frac{{\rm i}}{\hbar (2\pi)^4} \int d^4k \int d^4s \,\frac{\delta \left(\underleftrightarrow{S} \overleftrightarrow{\Phi}\right)}{\hat{\delta\Phi}^{\alpha}} \,\left[ A_{s,k}\, e^{{\rm i} \underleftrightarrow{K}.\overleftrightarrow{X}} e^{\frac{{\rm i}}{\hbar} \underleftrightarrow{S} \overleftrightarrow{\Phi}} \right.\nonumber \\
&-&\left. B^{\dagger}_{k,s} \, e^{-{\rm i}\underleftrightarrow{K}.\overleftrightarrow{X}} e^{-\frac{{\rm i}}{\hbar} \underleftrightarrow{S} \overleftrightarrow{\Phi}}\right],
\end{eqnarray}
where
\begin{equation}
\frac{\delta \left(\underleftrightarrow{S} \overleftrightarrow{\Phi}\right)}{\hat{\delta\Phi}^{\alpha}} = \hat{S}_{\alpha},
\end{equation}
such that $\bar{\gamma}^{\alpha}$ are the $4\times 4$ matrices that generate the hyperbolic background (Riemannian) space-time and comply with the Clifford algebra, $\hat{S}_{\alpha} = \frac{1}{2} s \bar{\gamma}_{\alpha}$, $\hat{K}_{\alpha} = \frac{1}{2} k \bar{\gamma}_{\alpha}$, $\hat{\Phi}^{\alpha} = \frac{1}{2}\phi \bar{\gamma}^{\alpha}$ and $\hat{X}^{\alpha} = \frac{1}{2} x \bar{\gamma}^{\alpha}$. We define the operatorial products
\begin{eqnarray}
\underleftrightarrow{K}.\overleftrightarrow{X} & = & \frac{1}{4} \left\{\hat{K}_{\alpha}, \hat{X}_{\beta}\right\} \left\{\bar{\gamma}^{\alpha}, \bar{\gamma}^{\beta}\right\} , \nonumber \\
 \underleftrightarrow{S} \overleftrightarrow{\Phi}&=& \frac{1}{4}\left\{ \hat{S}_{\mu}, \hat{\Phi}_{\nu} \right\} \left\{ \bar{\gamma}^{\mu} ,\bar{\gamma}^{\nu}\right\}-\frac{1}{4}\left[ \hat{S}_{\mu}, \hat{\Phi}_{\nu} \right] \left[ \bar{\gamma}^{\mu}, \bar{\gamma}^{\nu}\right],  \label{invariant}
\end{eqnarray}
such that in order for quantize the spin, we shall propose the universal invariant ($n$-integer):
\begin{equation}
\left<B\left|\underleftrightarrow{S} \overleftrightarrow{\Phi}\right|B\right> = s \phi \,
\mathbb{I}_{4\times 4} = (2\pi n \hbar) \,\mathbb{I}_{4\times 4}.
\end{equation}
In this work we shall deal only with bosons, in which the creation and destruction operators must comply\cite{MM}
\begin{eqnarray}
\frac{4 s^2 \,  L^2_p}{\hbar^2} \left(|A_{k,s}|^2 - |B_{k,s}|^2\right) &=& 0, \,\pm \left(\frac{c^3 M^3_p}{\hbar}\right)^2, \label{q22}
\end{eqnarray}
where $c$ is the speed of light, $M_p$ is the Planck mass and $h=2\pi \hbar$ the Planck constant. The conditions (\ref{q2}) are required for scalar bosons (the first equality) and vector, or
tensor bosons (the second equality). On the other hand, in order for the expectation value of the energy to be positive: $\left< B\left| {\cal H}\right|B \right> \geq 0$, we must choose the negative signature in the second equality of (\ref{q2}). Furthermore, conditions (\ref{q2}) are required for scalar bosons (the first equality) and vector, or
tensor bosons (the second equality). On the other hand, in order for the expectation value of the energy to be positive: $\left< B\left| {\cal H}\right|B \right> \geq 0$, we must choose the negative signature in the second equality of (\ref{q2}). The expectation value for the local particle-number operator for bosons with wave-number norm $k$ and spin $s$, $\hat{N}_{k,s}$, is given by\footnote{To connect the Fock-space theory and the ordinary quantum mechanics one can introduce the wave function in position space by using the definition of a kind of $n_{k,s}$-particle state vector that describes a system of $n_{k,s}$ particles that are localized in coordinate space at the points ${\bf x}_1; {\bf \phi}_1...{\bf x}_n; {\bf \phi}_n$:
\begin{displaymath}
\left|{\bf x}_1,{\bf x}_2,...,{\bf x}_n;{\bf \phi}_1,{\bf \phi}_2,...,{\bf \phi}_n \right> = \frac{1}{\sqrt{n_{k,s}!}} \hat{\slashed{\Psi}}^{\dagger}({\bf x}_1; {\bf \phi}_1)...\hat{\slashed{\Psi}}^{\dagger}({\bf x}_n; {\bf \phi}_n)\left|B\right>,
\end{displaymath}
where here $\left|B\right>$ is our reference state. This state is not a vacuum state because it describes a curved background state, but describes the Riemannian (classical) reference with respect to which we describe the quantum system.}
\begin{equation}\label{np}
\left<B\left|\hat{N}_{k,s}\right|B\right> =-n_{k,s}\,\left(\frac{\hbar}{c^3 M^3_p}\right)^2\,\int d^4x \sqrt{-g} \int d^4\phi \,\left< B\left| \left[\hat{\slashed{\Psi}}({\bf x}, {\bf \phi}), \hat{\slashed{\Psi}}^{\dagger}({\bf x}, {\bf \phi}) \right]\right|B \right>=n_{k,s} \,\mathbb{I}_{4\times4},
\end{equation}
where the slashed spinor fields are: $\hat{\slashed{\Psi}}=\bar{\gamma}^{\mu} {\hat\Psi}_{\mu}$, $\hat{\slashed{\Psi}}^{\dagger} = \left(\bar{\gamma}^{\mu} {\hat\Psi}_{\mu}\right)^{\dagger}$. Furthermore, these fields comply with the algebra
\begin{equation}
\left< B\left| \left[\hat{\slashed{\Psi}}({\bf x}, {\bf \phi}), \hat{\slashed{\Psi}}^{\dagger}({\bf x}', {\bf \phi}') \right]\right|B \right>
 = \frac{4 s^2 \,L^2_p}{\hbar^2} \left(|A_{k,s}|^2 - |B_{k,s}|^2\right) \,\,  \sqrt{\frac{\eta}{g}} \,\,\delta^{(4)} \left({\bf x} - {\bf x}'\right) \,\delta^{(4)} \left({\bf \phi} - {\bf \phi}'\right),
\end{equation}
which must be nonzero in order to particles can be created. Notice that this is the case for bosons with spin non-zero, but in the case of scalar bosons, which have zero spin, one obtains that $\left(|A_{k,s}|^2 - |B_{k,s}|^2\right)=0$, and $\left< B\left| \left[\hat{\slashed{\Psi}}({\bf x}, {\bf \phi}), \hat{\slashed{\Psi}}^{\dagger}({\bf x}', {\bf \phi}') \right]\right|B \right>=0$. This result is valid in any relativistic scenario.

As was demonstrated in a previous work, massless neutral bosons, with $\xi=0$, that describes the dynamics (\ref{a2}) with (\ref{q2}) and (\ref{q22}), and propagate in an arbitrary background space-time, must comply following equation for the wave-numbers of bosons\cite{SMM}:
\begin{eqnarray}
\left[\bar{\gamma}^{\beta},\, \bar{\gamma}^{\theta}\right]_{,\theta} &-& \frac{1}{2} g^{\beta\theta} \left(\bar{\gamma}^{\nu}\right)_{,\theta} \bar{\gamma}_{\nu}
-2{\rm i}\, k^{\beta} \mathbb{I}_{4\times4} + \frac{1}{2} g^{\nu\theta} \left(\bar{\gamma}^{\beta}\right)_{,\theta} \bar{\gamma}_{\nu} +\frac{{\rm i}}{2} \bar{\gamma}^{\beta} \underleftrightarrow{k} \nonumber \\
&= &  \frac{s^2}{2\hbar^2} \left\{ \begin{array}{cc}  \nu \, \\ \theta \, \nu  \end{array} \right\} \left[
\bar{\gamma}^{\theta}, \, \bar{\gamma}^{\beta} \right] + \frac{1}{2} g^{\beta\theta}  \left\{ \begin{array}{cc}  \mu \, \\ \nu \, \theta  \end{array} \right\} \bar{\gamma}^{\nu}
\bar{\gamma}_{\mu} - \frac{1}{2} g^{\mu \theta}  \left\{ \begin{array}{cc}  \beta \, \\ \nu \, \theta  \end{array} \right\} \bar{\gamma}^{\nu} \bar{\gamma}_{\mu}, \label{em}
\end{eqnarray}
where $\underleftrightarrow{k} = k^{\alpha} \bar{\gamma}_{\alpha}$ and $\bar{\gamma}_{\alpha}= E^{\mu}_{\alpha} \gamma_{\mu}$ are the components of the basis on the background metric, which are related by the vielbein $E^{\mu}_{\alpha}$ with the $4\times 4$ matrices $\gamma^{\mu}$ on the Minkowski spacetime. In our case we shall use cartesian coordinates to describe spacial coordinates. In this paper we shall use the Weyl representation of the $\gamma$-matrices to generate the hyperbolic space-time.

\section{Quantum action}

In order to describe the quantum action due to photons and gravitons, we shall propose
\begin{equation}\label{act1}
I_Q = \frac{c^4}{16 \pi G}  \int\, d^4x\,\int\, d^4\phi \,\sqrt{-g}\, \left<B\left| \bar{\gamma}^{\mu} \bar{\gamma}^{\nu} \left( \hat{\delta R}^{\alpha}_{\mu\nu\alpha} +
\hat{\delta R}^{\alpha}_{\alpha\mu\nu}\right) \right|B\right>,
\end{equation}
where
\begin{equation}
\bar{\gamma}^{\mu} \bar{\gamma}^{\nu} = \frac{1}{2} \left\{\bar{\gamma}^{\mu}, \bar{\gamma}^{\nu}\right\} - \frac{1}{2} \left[\bar{\gamma}^{\mu}, \bar{\gamma}^{\nu}\right].
\end{equation}
The expression (\ref{act1}) can be written as\cite{MM}
\begin{equation}\label{act2}
I_Q = \frac{c^4}{16 \pi G}  \int\, d^4x\,\int\, d^4\phi \,\sqrt{-g}\, \left<B\left| \frac{1}{2} \left\{\bar{\gamma}^{\mu}, \bar{\gamma}^{\nu}\right\} \hat{U}_{\mu\nu} -
\frac{1}{2} \left[\bar{\gamma}^{\mu}, \bar{\gamma}^{\nu}\right]\hat{V}_{\mu\nu}-\frac{1}{2} \left[\bar{\gamma}^{\mu}, \bar{\gamma}^{\nu}\right] \hat{\Sigma}_{\mu\nu} \right|B\right>,
\end{equation}
where the tensor $\hat{U}_{\mu\nu}$ is purely symmetric and the tensors $\hat{V}_{\mu\nu}$ and $\hat{\Sigma}_{\mu\nu}$ are purely antisymmetric. By defining the new antisymmetric tensors\cite{MM}
\begin{equation}
\hat{N}_{\mu\nu} = \frac{1}{2} \hat{V}_{\mu\nu} - \frac{1}{4} \hat{\Sigma}_{\mu\nu}, \qquad \hat{M}_{\mu\nu} = \frac{1}{2} \hat{V}_{\mu\nu} + \frac{1}{4} \hat{\Sigma}_{\mu\nu},
\end{equation}
se obtain that the quantum action (\ref{act2}) can be written as
\begin{equation}\label{act3}
I_Q = \frac{c^4}{16 \pi G}  \int\, d^4x\,\int\, d^4\phi \,\sqrt{-g}\, \left<B\left| \frac{1}{2} \left\{\bar{\gamma}^{\mu}, \bar{\gamma}^{\nu}\right\} \hat{U}_{\mu\nu} -
\frac{1}{2} \left[\bar{\gamma}^{\mu}, \bar{\gamma}^{\nu}\right] \left(3 \hat{M}_{\mu\nu}- \hat{N}_{\mu\nu} \right) \right|B\right>,
\end{equation}
that provides all the information about the geometrical quantum nature of the action. In this work we are interested only in the contributions due to photons and gravitons, which are given by the last term in (\ref{act3}). In particular we shall study this contribution in a power-law inflationary model.

\section{Power-law inflation}

We are interested in studying the emission of bosonic spinor field particles during inflation. In particular, we shall consider
the case of an inflationary universe where the scale factor of the universe describes a power-law expansion, and the line
element related with the background semi-Riemannian curvature, is
\begin{equation}
d\hat{S}^2 = \hat{g}_{\mu\nu} d\hat{x}^{\mu} d\hat{x}^{\nu}= d\hat{t}^2 - a^2(t) \hat{\eta}_{ij} d\hat{x}^i d\hat{x}^j,
\end{equation}
where the {\em hat} denotes that the metric tensor is defined over a semi-Riemannian manifold. We shall define the action ${\cal I}$
on this manifold, so that the background action describes the expansion driven by a scalar field, which is minimally coupled to gravity
\begin{equation}
{\cal I} = \int \, d^4x\, \sqrt{-\hat{g}}\, \left[ \frac{{\cal \hat{R}}}{2\kappa} +  \left[ \frac{1}{2}\dot\phi^2 - V(\phi)\right]\right],
\end{equation}
In power-law inflation the scale factor of the universe and the Hubble parameter, are given respectively by\cite{prd96}
\begin{equation}
a(t)= \beta \, t^p, \qquad H(t)= \frac{p}{t},
\end{equation}
where $\beta= {a_0 \over t^p_0}$, $a_0$ is the initial value of the scale factor, $t_0$ is the initial value of the cosmic time, and the background solution for the
inflaton field dynamical equation
\begin{equation}
\ddot\phi + 3 \frac{\dot{a}}{a} \dot\phi + V'(\phi) =0,
\end{equation}
is
\begin{equation}
\phi(t) = \phi_0\left[1 - {\bf ln} \left(\frac{\alpha }{4 \pi \phi^2_0\,G} t\right)\right],
\end{equation}
where $p = (\kappa/2) \phi^2_0$, $\beta = \frac{a_0}{t^p_0}$ and $\alpha = H_f$ is the value of the Hubble parameter at the end of inflation.
The scalar potential can be written in terms of the scalar field
\begin{equation}
V(\phi) = \frac{3}{\kappa H^2_f} \left(1 - \frac{2}{3 \kappa \phi^2_0}\right) e^{2 (\phi/\phi_0)},
\end{equation}
which decreases with $\phi \geq \phi_0$, such that $\phi_0$ is the value of the inflaton field when inflation starts.

In order to describe the dynamics of the spinor fields, we shall use a metric conformal to a Minkowski one
\begin{equation}
d\hat{S}^2 = a^2(\tau) \left[ d\hat{\tau}^2 -  \hat{\delta}_{ij} d\hat{x}^i d\hat{x}^j \right],
\end{equation}
where the conformal time, $\tau <0$, is related to the cosmic time $t$, as
\begin{equation}
\tau =  \frac{-c \,t_0^p}{t^{(p-1)} (p-1)}.
\end{equation}
Here, $c$ is the speed of light and the scale factor written in terms of the conformal time, is
\begin{equation}
a(\tau) = \frac{\left[(p-1) {t_0}^{p} \right]^{\frac{p}{p-1}} }{{t_0}^{p}} \left(\frac{-c}{\tau}\right)^{\frac{p}{p-1}},
\end{equation}
and the nonzero Levi-Civita connections, are
\begin{equation}\label{con}
4\left\{ \begin{array}{cc}  0 \, \\ 0 \, 0  \end{array} \right\} =4\left\{ \begin{array}{cc}  0 \, \\ 1 \, 1  \end{array} \right\} = 4\left\{ \begin{array}{cc}  0 \, \\ 2 \, 2  \end{array}
\right\} = 4\left\{ \begin{array}{cc}  0 \, \\ 3 \, 3   \end{array} \right\}=\left\{ \begin{array}{cc}  \theta \, \\ 0 \, \theta  \end{array} \right\} = \frac{-4p}{(p+1)} \frac{c}{\tau}.
\end{equation}

\subsection{DYNAMICS OF NEUTRAL MASSLESS, $s=\hbar$-BOSONS (PHOTONS) IN POWER-LAW INFLATION}

In this section we shall study the dynamics of arbitrary neutral massless bosons with unitary spin. We shall make use of the equation (\ref{em}), which in the case of $s=\hbar$ can be decomposed in two equations
\begin{eqnarray}
\left[\bar{\gamma}^{\beta},\, \bar{\gamma}^{\theta}\right]_{,\theta} & = &  \frac{s^2}{2\hbar^2} \left\{ \begin{array}{cc}  \nu \, \\ \theta \, \nu  \end{array} \right\} \left[
\bar{\gamma}^{\theta}, \, \bar{\gamma}^{\beta} \right], \label{f1} \\
\frac{1}{2} g^{\beta\theta} \left(\bar{\gamma}^{\nu}\right)_{,\theta} \bar{\gamma}_{\nu}
+2{\rm i}\, k^{\beta} \mathbb{I}_{4\times4} - \frac{1}{2} g^{\nu\theta} \left(\bar{\gamma}^{\beta}\right)_{,\theta} \bar{\gamma}_{\nu} -\frac{{\rm i}}{2} \bar{\gamma}^{\beta} \underleftrightarrow{k} &=& -\frac{1}{2} g^{\beta\theta}  \left\{ \begin{array}{cc}  \mu \, \\ \nu \, \theta  \end{array} \right\} \bar{\gamma}^{\nu}
\bar{\gamma}_{\mu} + \frac{1}{2} g^{\mu \theta}  \left\{ \begin{array}{cc}  \beta \, \\ \nu \, \theta  \end{array} \right\} \bar{\gamma}^{\nu} \bar{\gamma}_{\mu}. \nonumber \\
\label{f2}
\end{eqnarray}
In our case $\bar{\gamma}^{\mu}= E^{\mu}_a\,\gamma^a$, where the vielbein  are given by: $E^{\mu}_a= a^{-1}(\tau) \, \delta^{\mu}_a$. With the connections (\ref{con}), the equations (\ref{f1}) and (\ref{f2}), hold
\begin{eqnarray}
\left[\bar{\gamma}^{i},\, \bar{\gamma}^{0}\right]_{,0} & = &  \frac{1}{2} \left\{ \begin{array}{cc}  \nu \, \\ 0 \, \nu  \end{array} \right\} \left[
\bar{\gamma}^{0}, \, \bar{\gamma}^{i} \right], \label{ff1} \\
k^j \mathbb{I}_{4\times4} & = &  \frac{1}{6}
\left[\gamma^j,\gamma_{\alpha}\right] k^{\alpha}- \frac{i}{6} \frac{1}{a(\tau)} \frac{d\left[a^{-1}(\tau)\right]}{d\tau} \left[\gamma^j,\gamma_0\right], \label{ff2} \\
k^0 \mathbb{I}_{4\times4} & = &  \frac{1}{6}
\left[\gamma^0,\gamma_{\alpha}\right] k^{\alpha}. \label{f33}
\end{eqnarray}
where subscripts $i,j$ can take values from $1$ to $3$. The wave-numbers take the values
\begin{equation}
k^0= -\left(\frac{{\rm i}}{4}\right)\,f(\tau), \qquad k^1=  k^2=0, \qquad k^3= \pm  \,\left(\frac{3{\rm i}}{4}\right)\,f(\tau),
\end{equation}
where
\begin{equation}\label{f}
f(\tau)=-\frac{p\left[(p-1) t^p_0 c\right]^{\frac{2p}{p-1}} t^{-2p}_0}{(p-1)}\,(-\tau)^{\frac{p+1}{p-1}}.
\end{equation}
The squared-norm of massless $s=\hbar$-bosons on physical coordinates is
\begin{equation}
\frac{\left|k\right|^2}{a^2}= \frac{f^2(\tau)}{2} >0,
\end{equation}
which tends to zero, as $\left(-\tau\right)  \rightarrow 0$. The reason by which this value is nonzero is because we are describing massless particles that propagate in the $z$-direction in an isotropic universe which is accelerating expanding. However, due to the fact we are dealing with photons, the $k$-squared-norm must be null on physical coordinates. Therefore, the effective frequency and the $z$-component of the wave-number on physical coordinates should be altered in the following manner:
\begin{equation}
\left(\frac{\omega}{c}\right)^2 \equiv \left(\tilde{k}^0\right)^2 = \left[\Im\left(\frac{{k}^0}{a(\tau)}\right)\right]^2 + \frac{\left|k\right|^2}{a^2}=
\left(\frac{3}{4}\right)^2\,f^2(\tau), \qquad \left(\tilde{k}^3\right)^2 =\left(\frac{3}{4}\right)^2\,f^2(\tau),
\end{equation}
such that $g_{\alpha\beta}\,\, \tilde{k}^{\alpha} \tilde{k}^{\beta}=0$. Therefore, the redefined physical values $\tilde{k}^{\alpha}$, should be the values experimentally measured.

\subsection{DYNAMICS OF NEUTRAL MASSLESS, $s=2\hbar$-BOSONS (GRAVITONS) IN POWER-LAW INFLATION}

In order to study the emission of gravitons during power-law inflation, we shall use the equation (\ref{em}), for the case $s=2\hbar$. In this case such that equation can be decomposed in the equations
\begin{eqnarray}
\left[\bar{\gamma}^{\beta},\, \bar{\gamma}^{\theta}\right]_{,\theta} & = &  \frac{s^2}{8\hbar^2} \left\{ \begin{array}{cc}  \nu \, \\ \theta \, \nu  \end{array} \right\} \left[
\bar{\gamma}^{\theta}, \, \bar{\gamma}^{\beta} \right], \label{g1} \\
\frac{1}{2} g^{\beta\theta} \left(\bar{\gamma}^{\nu}\right)_{,\theta} \bar{\gamma}_{\nu}
+2{\rm i}\, k^{\beta} \mathbb{I}_{4\times4} - \frac{1}{2} g^{\nu\theta} \left(\bar{\gamma}^{\beta}\right)_{,\theta} \bar{\gamma}_{\nu} -\frac{{\rm i}}{2} \bar{\gamma}^{\beta} \underleftrightarrow{k} &=& -\frac{1}{2} g^{\beta\theta}  \left\{ \begin{array}{cc}  \mu \, \\ \nu \, \theta  \end{array} \right\} \bar{\gamma}^{\nu}
\bar{\gamma}_{\mu} + \frac{1}{2} g^{\mu \theta}  \left\{ \begin{array}{cc}  \beta \, \\ \nu \, \theta  \end{array} \right\} \bar{\gamma}^{\nu} \bar{\gamma}_{\mu} \nonumber \\
&+&\frac{3 \,s^2}{8\hbar^2} \left\{ \begin{array}{cc}  \nu \, \\ \theta \, \nu  \end{array} \right\} \left[
\bar{\gamma}^{\theta}, \, \bar{\gamma}^{\beta} \right]. \label{g2} \nonumber \\
\end{eqnarray}
Using the connections (\ref{con}), we obtain the wave-numbers values during power-law inflation
\begin{eqnarray}
k^j \mathbb{I}_{4\times4} & = &  \frac{1}{6}
\left[\gamma^j,\gamma_{\alpha}\right] k^{\alpha}-  \frac{{\rm i}}{6\,a(\tau)} \frac{d\left[a^{-1}(\tau)\right]}{d\tau} \left[\gamma^j,\gamma_0\right] +
\frac{4\,{\rm i}}{a(\tau)} \frac{d\left[a^{-1}(\tau)\right]}{d\tau} \left[\gamma^0,\gamma^j\right], \label{gg1} \\
k^0 \mathbb{I}_{4\times4} & = &  \frac{1}{6}
\left[\gamma^0,\gamma_{\alpha}\right] k^{\alpha}. \label{gg2}
\end{eqnarray}
The resulting wave-number values are
\begin{equation}
k^0= \left(\frac{23\,{\rm i}}{4}\right)\,f(\tau), \qquad k^1=  k^2=0, \qquad k^3= \mp  \, \left(\frac{69\,{\rm i}}{4}\right)\,f(\tau),
\end{equation}
with $f(\tau)$ given by (\ref{f}). On physical coordinates, the squared-norm for gravitons, is
\begin{equation}
\frac{\left|k\right|^2}{a^2}= \frac{529}{2}\,f^2(\tau) >0,
\end{equation}
which tends to zero, as $\left(-\tau\right)\rightarrow 0$. Here, we have the same problem that in the case of photons; the reason by which this value is nonzero is because we are describing particles that propagate in the $z$-direction in an isotropic universe which is accelerating expanding. Therefore, in order for the $k$-squared-norm be null, the frequency and $z$-wave number must be altered on physical coordinates, $\tilde{k}_{\alpha}\tilde{k}^{\alpha}\equiv g_{\alpha\beta} \,\,\tilde{k}^{\alpha} \tilde{k}^{\beta}=0$:
\begin{equation}
\left(\frac{\omega}{c}\right)^2 \equiv \left(\tilde{k}^0\right)^2 = \left[\Im\left(\frac{{k}^0}{a(\tau)}\right)\right]^2 + \frac{\left|k\right|^2}{a^2}=
\left(\frac{69}{4}\right)^2\,f^2(\tau),\qquad \left(\tilde{k}^3\right)^2 =\left(\frac{69}{4}\right)^2\,f^2(\tau).
\end{equation}
These should be the values measured in an experiment.

\subsection{Quantum action of photons and gravitons}

In order to calculate the quantum action (\ref{act3}) due to gravitons and photons, we must take into account the  expression (\ref{q2}). If we consider only massless bosons with spin $s=\hbar$ (photons), we
obtain
\begin{equation}
\left.I_Q\right|_{pho} = \frac{3}{64}\, h.
\end{equation}
On the other hand, for $s=2\hbar$-bosons (gravitons), the action (\ref{act3}) results to be
\begin{equation}
\left.I_Q\right|_{grav} = \frac{3}{16}\, h.
\end{equation}
Notice that in both cases the action is a fraction of the Planck constant $h$.

\section{Final comments}

We have studied the emission of photons and gravitons during power-law inflation, taking into account the recently introduced formalism for spinor fields. Our calculations show that
during inflation both kind of bosons were emitted with a frequency decreasing with the expansion of the universe as $H(\tau)/a(\tau)$. Therefore the wavelengths on physical coordinates related to these frequencies $\lambda_{Ph} = 2\pi/\omega \sim a/H$, increase, as expected, as the Hubble radius. If we consider that wavelengths at the beginning of inflation were of the order of the Planck length, at the end of inflation it should be at least $e^{60}$ times bigger than the Planck length.

As was shown in (\ref{np}), bosons with
$s=(1,2)\hbar$ can be created in any relativistic scenario, and
therefore can be created during inflation. Because the
wavelengths of photons and gravitons, are of the order of the
Hubble horizon, the present day frequency should be very low, of the order of
the inverse of the edge of the universe. However, they should be responsible for
very large-scale gravitational en electromagnetic primordial
fundamental wavelengths, which are coherent, and could be detected
in the future on the extreme (low-frequencies) range of the
primordial electromagnetic and gravitational spectrum. The source of these kinds of quantum radiation is in the last term of the action (\ref{act3}), which in the case of neutral massless bosons result to be a fraction of the Planck constant. This result is independent of the inflationary stage. In other words, this result is the same in any curved background. In the case of $s=\hbar$-bosons is $\left.I_Q\right|_{pho}=(3/64) \,h$, and for gravitons it is $\left.I_Q\right|_{grav}=(3/16) \,h$.

Our approach is something different than others recently propossed\cite{c1}\cite{c2} where gravitational waves appears as tensor fluctuations of the background metric tensor. In our case, an unified description of quantum spinor fields is developed and all these fields are vectorial. Therefore, photons and gravitons describe a vector (massless) field dynamics (\ref{a2}) on an extended manifold, and they are distinguished by their spin value. In our case, we have considered these fields as massless, but it is possible to describe massive photons and gravitons using $\xi \neq 0$ in the dynamics for bonsons[the reader can see \cite{MM}]. The existence of massive gravitons would be responsible for deviations of the Newton's gravitational potential, which lead to Yukawa-like corrections that were calculated in\cite{c2} for the bound on a graviton mass derived by LIGO-VIRGO, is\cite{ligo-virgo}: $m_g < 1.2 \times 10^{-12}\,eV$.

\end{document}